\documentstyle[12pt]{article}
\begin{document}
\newcommand{\be}{\begin{equation}}
\newcommand{\ee}{\end{equation}}
\parskip 10pt
\parindent 25pt
\begin{flushright}  OKHEP-94-11 \\ hep-ex 9406003 \end{flushright}
\title{A Neural Network for Locating\\ the Primary Vertex in\\
a Pixel Detector}
\vspace{2.0cm}
\author{R. Kantowski and  Caren Marzban \\ \\
Department of Physics and Astronomy\\
University of Oklahoma, Norman, OK  73019 U.S.A. }
\maketitle
\vspace{3.0cm}
\begin{abstract}

Using simulated collider data for $p+p\rightarrow 2{\rm Jets}\ $
interactions
in a 2-barrel pixel detector,
a neural network is trained to construct the coordinate of the
primary vertex to a high degree of accuracy.
Three other estimates of this coordinate are also
considered and compared to that of the neural network. It is
shown that the network can match the best of the traditional
estimates.

\end{abstract}

\pagebreak
\section{Introduction}

Artificial Neural Networks (ANNs) are increasingly gaining
attention
within the high energy physics community [1]. The interest emanates
from a variety of features common to all ANNs; in short, they are
fast and robust (see [2] and references therein). The former property
has justified
the use of ANNs as triggers, and the latter has made them useful in
off-line analyses as well. Most applications
have been concerned with the general problem of discrimination
between
an object of one type from an object of another type,
e.g., a quark jet from a gluon jet or W/Z decays from the QCD
background.
There are a relatively small number of applications
estimating analog quantities, such as the invariant
mass of hadronic jets [3], the slope of a track, or the coordinate
of a primary vertex in a drift chamber [4].
The accuracies of the measurements have
been typically low ($\sigma \sim$ millimeters) due to the low
resolution of
the detectors involved. In this article we probe the limits of
accuracy with
which an ANN can estimate the z-coordinate (position along the beam
line) of the primary
vertex of a  pp collision in a pixel
detector, where spatial resolution is of the order of tens of
microns.
We also compare our results with some traditional methods, exhibiting
the
high accuracy of the neural network's estimates. The method is not
restricted to use in pixel detectors, but can be applied to
any detector with a barrel geometry.

\section{Neural Networks}

There exists a plethora of statistical techniques for analyzing data,
and in combination with the variations emerging due to specialized
needs, one is faced with the difficulty of choosing the ``best"
method of analysis. At the same time, traditional  methods invariably
have inherent assumptions that limit their applicability. For
instance,
distributions are commonly assumed to be gaussian - an assumption
that
may easily be violated. Neural
Networks, on the other hand, are free of assumptions regarding the
$a$-$priori$ distribution of the data\footnote[1]
{It would be premature, however, to conclude that ANNs are a panacea;
see [5] for some
examples, in the context of discrimination, where traditional
methods outperform neural networks.}. They are
designed to extract existing patterns from noisy data. The procedure
involves training a network with a large sample of representative
data,
after which one exposes the network to data not included in the
training set with the aim of predicting the new outcomes.
Specifically, a feed-forward ANN has some number of input and output
nodes, characterizing respectively the independent and dependent
variables
of an underlying map which is to be learned by the network. There may
also be one or more
hidden layers with some number of nodes on each. The value $\sigma_j$
of
the j-th node (on the first hidden layer)
is given by $\sigma_j=f(\sum_i \omega_{ji} \tilde{\sigma}_i +
\theta_j)$, where
$\omega_{ji}$ are the weights connecting the i-th input node (whose
value is $\tilde{\sigma}_i$) to the
j-th hidden node whose ``activation" threshold is $\theta_j $. A
similar rule
applies to the nodes on the second hidden layer, as well as
those on the output layer, with the values of the nodes on any layer
being
determined from the ones on the previous layer. Typically, the
activation
function $f$ for the first layer is a sigmoid function, and for any
remaining layer is either a sigmoid or a linear function.

Training an ANN involves minimizing the sum of the
differences-squared
of the
outputs of the network and the targeted values; the initial randomly
assigned weights are modified according to some learning rule in
order
to minimize this quantity (called the energy function). Subsequently,
the input nodes are assigned
values (not in the training set), and the value of the
output nodes, as determined from the trained weights leading to them,
are taken
to be the predicted value of the dependent variables in question. The
performance of the network is monitored by a validation set, which is
another set of known independent/dependent variables, not used in the
training,
and whose target values are
compared to the values predicted by the trained network; the
comparison
is done in a variety of ways, one of which will be discussed below.

The particular ANN program used in this study was a modified version
of
one obtained from [6]. The original source codes, written
in C++, were designed to be compiled and executed on DOS machines.
For our purposes the source codes were modified to run in the UNIX
environment. The modified version allows us to use a large number of
nodes on each layer as well as to view the network's weights. Usually
the weights are uninterpretable due to the presense
of hidden layers and the nonlinearity of the activation function;
however, in
the present study
the particular architecture of the network does allow for an
unambiguous interpretation. The activation function for all the
layers
is taken to be the logistic (or fermi) function,
$f(x)=1/(1+\exp(-x))$.

The energy surface whose minimum is to be found is well-known for
being
infested with local minima. The particular ANN program used here
offers two
methods for eluding the local minima - Simulated Annealing and a
Genetic
Algorithm. For this study we employed the former, both for initiating
a
set of weights that could then be evolved according to the learning
rule,
and for attempting to escape the local minimum when the learning rule
was
incapable of doing so. The particular learning rule adopted here was
the conjugate gradient method, a variation of the more familiar
back-propagation method [6].

In training an ANN some transformation of the data is inevitable. For
instance, given the range of the fermi function, one must scale the
target values to lie in the range 0 to +1 (for numerical
reasons, it is advisable to shrink that range to 0.1 to 0.9).
Because of the asymptotic behavior of the fermi function for both
small
and large values of the domain, it is beneficial to scale the
independent
variables to lie in a similar range as well. Later, we will discuss
one additional
transformation of the data.

\section{Vertex Detectors}

For the application at hand, we trained
a network using data simulating the reaction $p+p\rightarrow 2{\rm
Jets}\
+ X$
as would be detected by the SDC hybrid pixel detector
[7] being designed for use at the SSC until the project was
discontinued.
The geometry of the detector is that of eight concentric barrels
surrounding
the beam pipe. Only the inner two are covered with pixels wafers and
it is
data from these barrels we find sufficient to obtain the desired
resolution of the primary vertex's position ($z_0$) along the
colliding beam path.
The radii of the two barrels are $r_1=6$cm and $r_2=8$cm; however,
due to
an overlapping of adjacent pixel wafers (approximately 1 cm$^2$ in
size) attached to these barrels,
the radial coordinates of a particle (when it hits the pixels) differ
slightly from these values. Each wafer contains 12 columns by 64 rows
of 50 $\mu$m by 300 $\mu$m pixels and the length of the
barrels is approximately $34$cm ($z=-17$cm to $z=+17$cm).

The simulation program SDCSIM, modified to include the pixel detector
for both simulation and reconstruction, provided \footnote[2]{Phil
Gutierrez and Hong Wang of the OUHEP group provided this data for
us.}  a pair of
coordinates, ($z_1, r_1,\phi_1$) and  ($z_2, r_2,\phi_2$) for each
particle as it penetrated the two barrels. These hits were given by
the generator rather than by reconstruction; however,
we know from other simulation work that hits on the 2 barrels can
easily be correlated via a $\Delta \phi$ constraint
to identify a given particle). The true coordinate of the primary
vertex, $z_0$,
for each event was also provided by the simulation.  Because the
simulated SSC proton beams were narrow (5$\mu$m) and since we were
interested only in an accurate estimate of the
position of the primary vertex, the angular coordinates were ignored.
There are,  on average, 100  particles produced in high $p_t$ events
at
SSC (20 TeV on 20 TeV) energies and each could be used to estimate
$z_0$
by extrapolating a straight line in the z-r plane through the two
barrel
hits, back to the beam axis ($ r = 0$). For the $i$-th particle the
estimate is
\begin{equation}
z(i)={z_1(i)r_2(i)-z_2(i)r_1(i)\over r_2(i)-r_1(i)}.
\end{equation}

The goal was to estimate the true primary vertex coordinate $z_0$
from the distribution of $z(i)$ values. In addition to the
neural network, we considered 3 other
estimates - the mean, the point where the histogram maximum occurs,
and the
median of the distribution. Figure 1 shows the $z(i)$ distribution
for
a typical event, along with 3 estimates of the primary vertex.
Due to outliers the mean does  not give a very
good estimate of $z_0$. In these examples the  point where the peak
of
the distribution occurs gives a much better estimate of $z_0$.
However,
this result obviously dependents on the histogram's bin-size and can
only
be used for events that produce a unique maximum. The median is a
more
robust estimate, in that it is well-defined
and it is insensitive to the presense of outliers. A quantitative
measure of exactly how good these estimates are, along with that of
the neural network, will be presented below.

\section{Neural Networks in Vertex Detectors}

The neural network ultimately arrived at in this work was constructed
in two phases. In the first phase, a training set of 300 simulated
high $p_t$ pp events was used to find the optimal configuration
(including the number of input nodes) of the network. The second
phase adopted this configuration, but employed 800 simulated high
$p_t$ pp events as the training set, and it is this set that was
used for the remainder of the analysis.

The $z_0$ coordinates of the primary vertices were scaled to lie in
the range 0.1 to 0.9 and the single output node of the network was
assigned this quantity ($\tilde{z}_0\equiv 0.4 z_0/17 + 0.5$).
Since the number of outgoing particle tracks per collision is
variable, we
randomly chose a fixed number (55) from  each event and assigned the
corresponding \underbar{scaled} $z(i)$ (i.e., $\tilde{z}(i)$) values
to
the 55 input nodes.
The number 55 was chosen in the first phase
to maximize the performance of the network (See Figure 2). In the
first phase
282 events had 55 or more tracks and could be used, whereas in the
second phase, 766 could be used.

The same validation set containing 100 events was used in both
phases.
The number of events in this set having 55 or more
tracks was 94.

A network with no hidden layers, when presented the randomly selected
data
(tracks), did not meet the required degree of accuracy.
Figures 3a and 3b, show the actual $z_0$ coordinate versus the
predicted value,
when a trained network is exposed to the training set of 300 events
(2a),
as well as to the validation set (2b). For perfect training and
perfect
prediction we would expect all points to lie on a straight line
of slope one (for both plots). It is clear that this network is not
performing well at all, although partial learning and prediction
is clearly present in both plots.
Larger networks with one and two hidden layers and
with a variety of nodes on each layer, were tested with similar
results. The possibility that the poor performance of these networks
was
due to the algorithm being trapped in a (shallow)
local minimum was ruled out upon testing a host of parameters
that control the simulated annealing.

It is well-known in neural net circles that an appropriate
representation of the training data is crucial for the proper
training,
as well as for the predictive uses of the network. In this case,
the inability of the network to learn and to predict can be traced,
not to any shortcoming of the network itself, but to the data that
it has been expected to learn. Since the 55 tracks are picked
randomly, and then presented to the network as input, a given
input node is assigned a random value as the network proceeds from
one event to the next. The weight connected to a given input
node is then being updated according to a gradient rule with
no unique direction for the gradient itself.

To resolve this problem, we ordered the 55 input values before
presenting
them to the network. This gives each input node a unique ``identity"
and leads to a unique direction for the gradient updating and
considerably improves the performance of the network. However,
if we preprocess the training set in one additional way, not only
does
the network's  performance improve, but a
simple interpretation of exactly what the network is ``doing" can be
found -
an attribute whose absence is often considered a disadvantage
of neural networks. In particular, we train the network, not
simply using the scaled coordinate $\tilde{z}_0$ of the primary
vertex
as the output node, but with a further transformed coordinate
$\hat{z}_0\equiv f(\tilde{z}_0)$, where $f$ is the
logistic function, given above.
Of course, this choice is motivated by the activation function
itself being the logistic function\footnote[3]{This use of the
logistic function in both the target value and as an activation
function, may seem equivalent to using a linear activation
function and using $\tilde{z}_0$ itself as the target value. However,
that networks with sigmoid activation functions have far better
convergence properties, precludes a consideration of that
alternative.}.
To take advantage of the 0-1 range limit of the logistic function,
this time we scale the data (input and output) according to
$\tilde{z}_0\equiv 0.9 z_0/17$.

Figures 4a, 4b, and 5a, 5b show the performance, on both the training
set and
the validation set, of a network trained with 10 events and
766 events, respectively. It is now clear that the network is both
learning
and predicting to a high degree of accuracy. That the points in
Figure
5a lie on a straight line of slope one, may suggest that the
network is overtrained. However, a counting of the free parameters of
the network shows that no overfitting is occurring; the number of
data points (766) is larger than the number of free parameters (56).
A network with
no hidden layers, 55 input nodes, and 1 output node has 55 weights
and
1 threshold.
It is the absence of a hidden layer that makes possible the
interpretation
of the weights (see below).

To gain a quantitative measure of the accuracies involved, we compute
\begin{equation}
 E = \sqrt{ \frac{1}{N} \sum_{n=1}^{N} (o_n-t_n)^2}\;\;,
\end{equation}
as an estimate of the error, where $o_n$ and $t_n$ refer to the
output
of the network and the
target value, respectively, of the n-th event. $N$ is the total
number of events in either the training set or the validation set,
depending on which error is being reported; we shall use the same
symbol, $E$, for both cases. It is the transformed variables
$\hat{z}$ which we use to calculate an $\hat E$, and then scaled by
$E =  \hat E /(0.013)$ to present here. This scaling gives $E$ the
interpretation of a standard deviation of the distribution of the
\underline{unscaled} $z$'s.
For comparison $E$ is the appropriate
measure of error to use for the other 3 estimates (the mean, the
histogram
maximum, and the median) as well.
Noting that $E$ is the standard
deviation of a distribution of the
residuals $(o_n-t_n)$, we plot a histogram of the residuals for both
the training set and the validation set (Figures 6a and 6b).
Figure 7 shows
the error $E$, in microns, as a function of the number of
events in the validation set. Evidently,
the median of the distribution is the best of the three, reaching
$E=70\mu$.

Figure 8 plots $E$ of the validation set (94 events) as a
function of the number of events in the training set.
It is evident that the network is predicting the target value
with increasing accuracy, as the training set is enlarged.
This general behavior is typical of all ANNs, and it can be
shown [8] that it falls off as $\sim 1/N$, approaching a limit whose
value depends on the architecture of the network. Upon
training with 766 events, $E$ of the validation
set is at 89 microns. At this point the network has
outperformed the mean and the maximum of the distribution,
and the median is the only reasonable contender.

That the mean is not a good estimate can be explained by the presence
of outliers that do exist in the distribution (not shown in Figure
5).
The median, on the other hand, is insensitive to the presence of
outliers,
hence it's superior performance. The maximum is also insensitive to
outliers,
but it suffers from ambiguities
arising from the bin size, such as multiple maxima. We also tested
an average over 5 bins around a maximum, and obtained only a minor
improvement.

Due to the size of the pixels, there is an error in $E$
that can be calculated from
\[\sigma_E= \frac{150 \mu}{\sqrt{12}} \times
\frac{\sqrt{1+(r_1/r_2)^2}}{1-(r_1/r_2)} \times \frac{1}{\sqrt{55}}
\times
\frac{1}{\sqrt{N}}\;\;.\]
The origins of the terms in this equation are as follows: The first
is due to the length of the pixel in the z direction, the second is
from the propagation of errors in equation (1), the third term is
due to 55 measurements per event, and the last term
comes from the propagation of errors through equation (2). With
$r_1/r_2=6/8$ we obtain $\sigma_E=3$ microns. This
suggests that the estimates for $z_0$ from the network and the other
3
measures do not overlap due to the intrinsic resolution of the
detector.

Given the simple architecture of the network, and the transformations
performed on the training data, it is possible to decipher what
the network is doing. Figure 9a shows the 55 weights connecting
each of the input nodes to the single output node, for a  network
trained
with non-ordered input values. This pattern resembles an
untrained network, thereby explaining the poor performance of the
network before the ordering. Figure 9b shows the weights for a
network
trained with ordered input values. Evidently,
the network is automatically  performing a cut (in z) and averaging
over the
remainder of the distribution. Of course, this is precisely
the procedure traditionally attempted, though in this case the
value of the cut is being optimally determined by the network,
as it minimizes the output error in the training process.

As a final point (of curiosity), it is interesting that the
network tends to under-weight the tracks immediately adjacent
to the central plateau. In order to determine if this
was a peculiarity of the particular network at hand, we trained
networks
using a varying number of input nodes (i.e. tracks), and found
that this behavior is persistent. The ``reason" for this diffractive
behavior
is unclear to us, apart from the fact that it is necessary to
optimize
the performance of the network.

{\bf Acknowledgements}

We are grateful to Andy Feldt and John Kuehler for assistance in
the use of C and Fortran. George Kalbfleisch and Eric Smith are
acknowledged for useful discussions. Also, we thank
Pat Skubic and Phil Gutierrez for providing the initial stimulus for
undertaking this research project.  This work was supported by the
Department of Energy, and the Southern Association for High Energy
Physics (SAHEP) funded by the Texas National Research Laboratory
Commission (TNRLC).

\end{document}